%% file: MaxLik_PRL.tex
\documentclass[ prl,
reprint,
 amsmath,amssymb,
 aps,
]{revtex4-2}

\usepackage{graphicx}
\usepackage{tikz}
\usepackage{dcolumn}
\usepackage{bm}
\usepackage{braket}

\begin{document}


\title{ Resolving the phase space  }

\author{Zden\v{e}k Hradil  }
\email{hradil@optics.upol.cz} 
\affiliation{Department of Optics, Palack\' y University, 17.
listopadu 12,  779~00 Olomouc, Czech Republic}
\author{Jaroslav \v{R}eh\'{a}\v{c}ek}
\affiliation{Department of Optics, Palack\' y University, 17.
listopadu 12,  779~00 Olomouc, Czech Republic}


\date{\today}


\begin{abstract}
We show that the effective resolution of quantum inference is determined by a sampling Gram operator that defines the experimentally accessible degrees of freedom and the reconstruction bandwidth supported by the data. Acting analogously to a transfer function in classical imaging, this operator together with scaling of deviance provides an operational criterion for distinguishing genuinely resolved features from structures arising from incomplete sampling, statistical fluctuations, or reconstruction assumptions. Reconstruction in the Gram eigenbasis emerges naturally as a measurement-adapted compression of the inference problem, retaining only those modes supported by typical measurement outcomes. Within finite frame theory, the same structure appears as the frame operator governing stable reconstruction. The resulting framework connects quantum tomography, inverse reconstruction, and measurement resolution, providing a practical tool for assessing whether fine phase-space structures inferred in contemporary experiments are genuinely resolved by the measurement or merely reconstructed by the model.

\end{abstract}

\maketitle



{\em Introduction: }  Information content of a measurement lies at the heart of scientific inference. Its importance extends far beyond quantum physics and was perhaps captured most vividly by Eddington’s famous Fishing Net Parable\cite{Eddington1939}. In Eddington’s analogy, an ichthyologist explores marine life using a fishing net with meshes two inches wide and concludes that no sea creature shorter than two inches exists. Such a conclusion, however, reflects the properties of the net rather than the properties of the sea. Before drawing conclusions about the observed world, one must first understand the resolving power of the measuring device.

The same principle applies to quantum measurements. Before attributing a particular feature to a quantum system, one should establish whether the measurement apparatus possesses sufficient resolution to detect that feature in the first place. This raises a fundamental question of quantum inference: which properties of a quantum state are genuinely supported by the observed data, and which remain inaccessible because they lie beyond the information capacity of the measurement?
We formulate this question within a general framework of quantum measurement and statistical inference. Quantum tomography serves as a paradigmatic example connecting detection and information since  more than three decades it  has been a central tool for characterizing quantum systems. Although the quantum state is not directly observable, it can be inferred from suitably chosen measurement data, leading to an estimated density operator consistent with experimental observations. As experimental capabilities have advanced, tomography has evolved from proof-of-principle demonstrations into a quantitative diagnostic tool capable of probing increasingly subtle structures of quantum states.

The modern framework of quantum tomography emerged from optical homodyne detection and reconstruction techniques based on inverse Radon transforms \cite{Vogel_89}, followed by the first experimental demonstrations \cite{Smithey1993}; for a comprehensive review, see Ref.~\cite{Lvovsky09}. Because these reconstruction procedures were linear, however, they did not guarantee physically valid density operators, motivating the development of statistical methods based on maximum-likelihood (MaxLik) estimation \cite{Hradil_1997,Hradil2004}. MaxLik tomography enforces positivity and has therefore become the standard approach in experimental quantum state reconstruction \cite{James_01,Lvovsky_2004}. A modern overview of quantum state estimation, encompassing Bayesian inference, statistical interpretation, and maximum-likelihood methods, is provided in Ref.~\cite{Englert2025}.

Despite the success of statistically motivated methods, a fundamental question remains largely unresolved: when experimental resources are limited, which features of a reconstructed quantum state are genuinely supported by the data, and which arise primarily from reconstruction assumptions? The question has become particularly important in contemporary experiments investigating highly structured nonclassical states, as exemplified by bosonic encodings, Schr\"{o}dinger-cat states \cite{Ourjoumtsev2006}, Gottesman--Kitaev--Preskill (GKP) states \cite{science24}, large-photon-number states \cite{Hlousek_19}, or sub-Planck phase-space structures \cite{moore2026}, where the achievable resolution of quantum tomography is often the limiting factor. Yet this distinction is frequently obscured because state reconstruction is typically performed within parametrized model spaces whose relation to the actual information content of the measurements remains implicit. While sophisticated statistical methods are available, experimental practice commonly relies on point estimators such as MaxLik reconstruction \cite{Ourjoumtsev2006,science24}, yielding a single density operator but providing little indication of which of its features are statistically well constrained by the data.

In this Letter we show that MaxLik quantum tomography implicitly defines a sampling Gram operator $G$  that quantifies the information acquired by the measurement and thereby determines which features of a quantum state are experimentally resolvable. The Gram operator plays the role of Eddington's fishing net: its eigenmodes determine which structures can be faithfully resolved and which remain fundamentally inaccessible.

This viewpoint naturally leads to a conservative reconstruction strategy in which the density operator is represented in the eigenbasis of the sampling operator, providing a physically transparent criterion for distinguishing experimentally supported features from artefacts arising from insufficient sampling or model assumptions. Proper normalization of the likelihood and a parametrization consistent with the measurement operator are essential for this task. Approaches that do not respect this structure  may generate apparent fine-scale features that are not warranted by the information content of the data, a risk inherent to several commonly used reconstruction procedures  \cite{James_01,Lvovsky_2004}.

{\em MaxLik tomography : }
The basic structure of MaxLik tomography relevant for the present analysis is briefly reviewed below in several steps. Consider a generic detection scheme in which a quantum state $\rho$ is probed by POVM elements $\{\Pi_i\}$, where the $i$th outcome is registered $n_i$ times. The corresponding probabilities are
\begin{align}
p_i(\rho)=\mathrm{Tr}(\rho \Pi_i),
\qquad
\sum_i \Pi_i = G \ge 0 .
\end{align}
In general the measurement is incomplete, since some detection channels may be absent. The measured statistics therefore defines a conditional multidimensional distribution with log-likelihood functional
\begin{align}
\label{nonconvex}
\log \mathcal{L}(\rho)
=
\sum_i n_i
\log
\left[
\frac{p_i(\rho)}
{\sum_k p_k(\rho)}
\right].
\end{align}

Maximization under the constraint $\mathrm{Tr}(\rho)=1$ yields the nonlinear extremal equation
\begin{align}
R(\rho)\rho = G\rho,
\label{maxlik_eq}
\end{align}
where
\begin{align}
R(\rho)
= g  \sum_i
\frac{f_i}{p_i(\rho)}
\Pi_i,\\
f_i=\frac{n_i}{ N}, \quad g = \sum p_k(\rho),    \quad N=  \sum_k n_k.
\end{align}
Introducing the rescaled operators
\begin{align}
R_G = G^{-1/2}RG^{-1/2},
\qquad
\rho_G = G^{1/2}\rho G^{1/2},
\end{align}
Eq.~(\ref{maxlik_eq}) takes the normalized form
\begin{align}
R_G\rho_G=\rho_G .
\end{align}

The transformation maps the incomplete measurement onto an effective complete measurement on the support of $G$, since
\begin{align}
\sum_i
G^{-1/2}\Pi_iG^{-1/2}
=
\mathbb{I}_G .
\end{align}

Due to this rescaling, the interpretation can be reduced to the case $G=\mathbb{I}$. The MaxLik solution may then be formally written as the completeness relation for renormalized POVM elements,
\begin{align}
\sum_i \Pi_i' = \mathbb{I},
\end{align}
with
\begin{align}
\Pi_i' =
\frac{f_i}{p_i(\rho)}\,\Pi_i,
\qquad
{\rm Tr}(\rho \Pi_i') \equiv f_i .
\end{align}
The Born rule is thus satisfied identically, while the completeness relation emerges self-consistently through the MaxLik iteration. This highlights the particular role of MaxLik estimation within quantum theory, where statistical consistency and quantum-mechanical structure become directly linked. Notice however,  that conditional log-likelihood \ref{nonconvex} is non-convex functional, but be treated as one parametric family of convex problems, see Supplementary Material for more details.

The nonlinearity of MaxLik tomography has a clear statistical origin. Different projections do not fluctuate equally: projections aligned with the true state yield nearly deterministic outcomes, whereas orthogonal projections exhibit substantially larger statistical fluctuations. Consequently, the measurement records cannot be treated with equal statistical confidence. The nonlinear MaxLik iteration therefore adaptively reweights the measurement operators according to their statistical relevance, effectively estimating both the quantum state and the confidence associated with individual measurement outcomes simultaneously.

The situation changes qualitatively in the presence of limited experimental resources, where the measurement probes the Hilbert space only incompletely. In this regime, the operator $G$ naturally quantifies the portion of Hilbert space effectively accessible to the detection process. It is therefore natural to use its eigenbasis
\begin{align}
G|g_k\rangle=\lambda_k|g_k\rangle
\end{align}
for the modal decomposition of the reconstructed density operator.
For projective measurements with POVM elements
\begin{align}
\Pi_i = |y_i\rangle\langle y_i|,
\end{align}
the operator $G$ is unitarily equivalent to the corresponding Gram matrix
\begin{align}
G_{ij}=\langle y_i|y_j\rangle ,
\end{align}
associated with the generally nonorthogonal measurement vectors $|y_i\rangle$.
For this reason we refer to $G$ as the Gram operator. Physically, it plays a role closely analogous to a transfer function in classical imaging, or to the effective number of degrees of freedom introduced by Toraldo di Francia more than half a century ago~\cite{ToraldodiFrancia:69}. The eigenvalues $\lambda_k$ determine how efficiently individual Hilbert-space modes are transmitted through the measurement process. Modes associated with large eigenvalues are reliably resolved, whereas modes corresponding to small eigenvalues are only weakly accessible and therefore highly sensitive to statistical noise.
 Quantum tomography may therefore be viewed as an imaging channel in Hilbert space, where the Gram operator $G$ defines the effective transfer bandwidth and the eigenvalues $\lambda_k$ determine the transmission efficiency of individual modes. 

Interestingly, the present framework also admits a natural formulation within finite frame theory \cite{Waldron2018}, where signal representation and reconstruction are described in terms of generally nonorthogonal and overcomplete bases. This connection provides additional insight into the role of the Gram operator and is discussed in more detail in the Supplementary Material. 

Besides the Gram operator acting in Hilbert space, the formalism naturally introduces the operator-space Gram matrix
\begin{align}
Q_{ij}=|\langle y_i|y_j\rangle|^2 ,
\end{align}
which characterizes the conditioning and noise sensitivity of the reconstruction map. The two structures play complementary roles: the Gram operator in state space determines the effectively accessible region of Hilbert space, whereas the Gram matrix in operator space quantifies the stability and statistical sensitivity of the inversion.

Operationally, if the likelihood functional is not consistently normalized and parametrized, as in some widely used practical implementations of MaxLik tomography~\cite{James_01,Lvovsky_2004}, the direct connection to the underlying quantum-mechanical structure, namely the simultaneous interplay between the completeness relation and the Born rule, becomes obscured. Although the resulting reconstructions may still reproduce the measured data well, particularly in overdetermined finite-dimensional settings where the number of detection events exceeds the number of free parameters, the associated Gram-operator structure remains implicit.
As a consequence, the effectively accessible region of Hilbert space is no longer determined operationally by the measurement itself, but must instead be introduced through an external or ad hoc truncation of the reconstruction space.

The central task of the present framework is to identify the subspace of Hilbert space in which the reconstructed state is genuinely supported by the available data. If the experiment provides $N$ distinct measurement projectors, the formally accessible space is given by their linear span ${\cal H}_{\mathrm{lin}}$. However, tomography within this space is not generally tomographically complete, since reconstruction of a density operator requires determination of both diagonal and off-diagonal matrix elements.

MaxLik tomography effectively reduces the reconstruction problem to the support of the Gram operator ${\cal H}_G$. Even within this accessible subspace, however, the measurement sensitivity is highly nonuniform. This nonuniformity is quantified by the eigenvalue spectrum of the Gram operator: modes associated with small eigenvalues correspond to directions in Hilbert space that are only weakly constrained by the data and are therefore highly sensitive to statistical fluctuations.

For purposes of quantum diagnostics, it is therefore natural to restrict attention to the subspace spanned by eigenmodes with sufficiently large eigenvalues, which defines the region where the measurement provides statistically robust information. 
 The modal decomposition defined by the eigenvectors of $G$ provides a physically meaningful basis for reconstruction, while the corresponding eigenvalues quantify how reliably individual modes are resolved. Its application to tomography provides a dual benefit: efficient low-dimensional representation of the target state together with suppression of weakly supported noisy modes.

The eigenvalue spectrum of the Gram operator provides a priori information about the sampling strength of individual reconstruction modes. The practical impact of this modal structure can be assessed directly from the measured data through the behaviour of the likelihood deviance as a function of reconstruction dimension.
Deviance (or equivalently the Kullback–Leibler divergence)  between the detected frequencies and the probabilities is defined as 
\begin{align}
D = \sum_k f_k \log f_k
+ \sum_k f_k \log \left(\sum_i p_i\right)
- \sum_k f_k \log p_k(\rho),
\end{align}
and  differs from the negative log-likelihood only by an additive constant. Small values of the deviance indicate a better agreement between the measured frequencies and the fitted probabilities. In this respect, the deviance plays a role analogous to the more familiar $ \chi^2$  statistic, which may be regarded as its Gaussian approximation.

As illustrated below, the dependence of the deviance on the reconstruction dimension in the Gram eigenbasis provides a direct operational criterion for determining the effective resolution of the measurement, or, in Eddington’s terminology, for determining the mesh size of the fishing net. The deviance exhibits a rapid decrease when the dominant Gram modes are included, indicating that additional dimensions contribute significantly to the description of the measured data. Beyond a certain dimension, the decrease becomes substantially weaker, signalling that newly added modes provide only marginal improvement of the reconstruction.
 Features contained within the dominant modes may be regarded as resolved, whereas structures requiring weakly sampled modes should be interpreted cautiously and ideally verified by additional measurements or more advanced statistical analysis.
Analogously to  spectral analysis in classical imaging reconstruction in the Gram eigenbasis driven by experimental data acts simultaneously as dimensional reduction and statistical filtering: it enables efficient low-dimensional parametrization while preventing unsupported structures from entering the reconstruction.

 Consequently, features recovered only after introducing weakly sampled modes may reflect reconstruction assumptions rather than experimentally supported information.
From this perspective, the commonly pursued goal of obtaining the ``best possible'' reconstruction may need to be reconsidered in favor of reconstructions whose effective resolution is demonstrably supported by the measurement statistics. This issue becomes increasingly important in contemporary experiments investigating bosonic encodings, GKP states, and other highly structured nonclassical states exhibiting fine phase-space features. Similar considerations are likely relevant beyond quantum-state tomography, including inverse reconstruction methods used in ultrafast optical pulse characterization and FROG-type spectrogram inversion \cite{Trebino2002,Seifert2016,Bhattacharjee2025}.


\begin{figure}[ht]
\centering
\includegraphics[width=0.48\columnwidth]{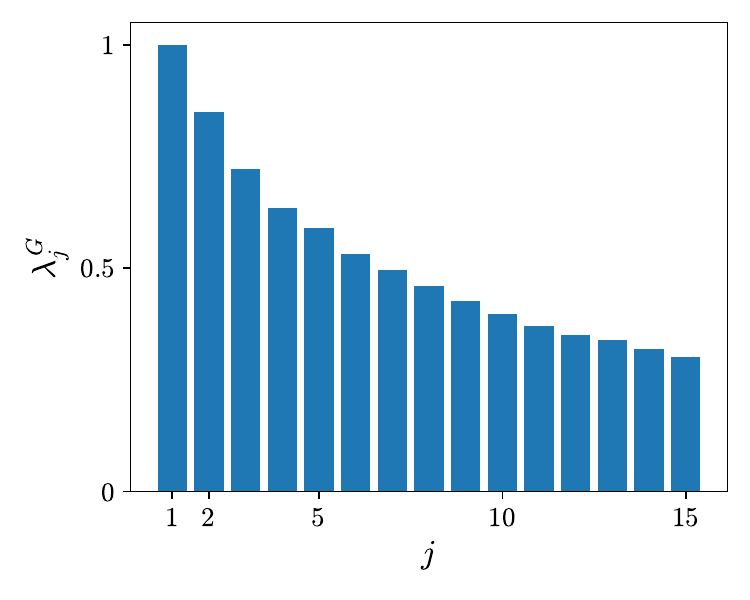}
\hfill
\includegraphics[width=0.48\columnwidth]{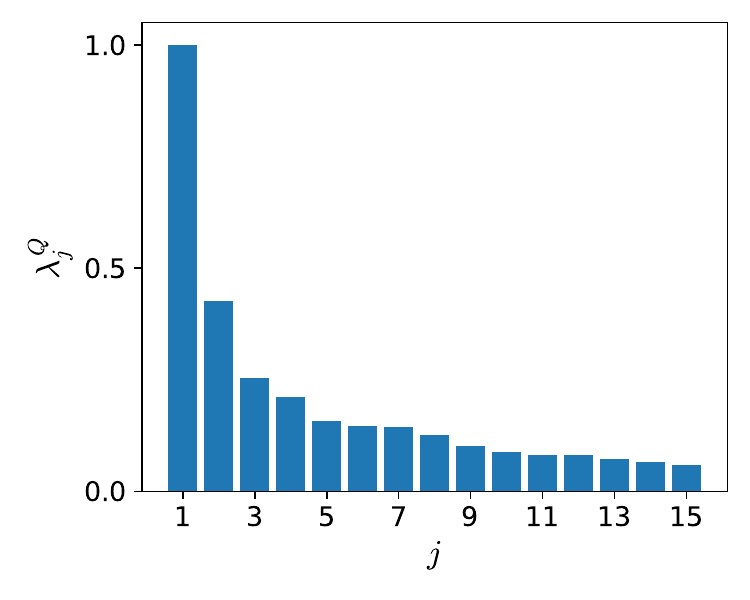}
\caption{
Eigenvalue spectra of the Gram operator $G$ (left) and the operator-space Gram matrix $Q$ (right) for homodyne tomography of a cat state. The spectrum of $G$ determines the effective reconstruction bandwidth, while $Q$ governs conditioning and noise sensitivity.
}
\label{Gram}
\end{figure}

\begin{figure}[ht]
\centering
\includegraphics[width=0.3\columnwidth]{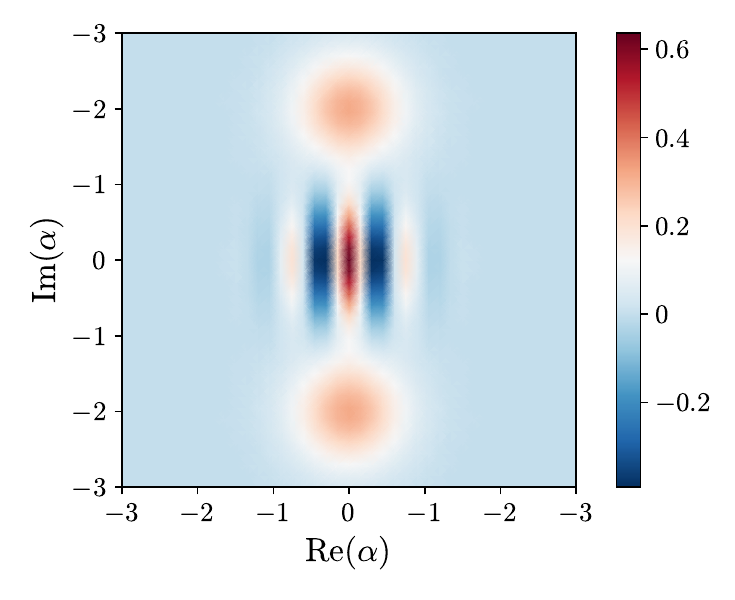}
\includegraphics[width=0.3\columnwidth]{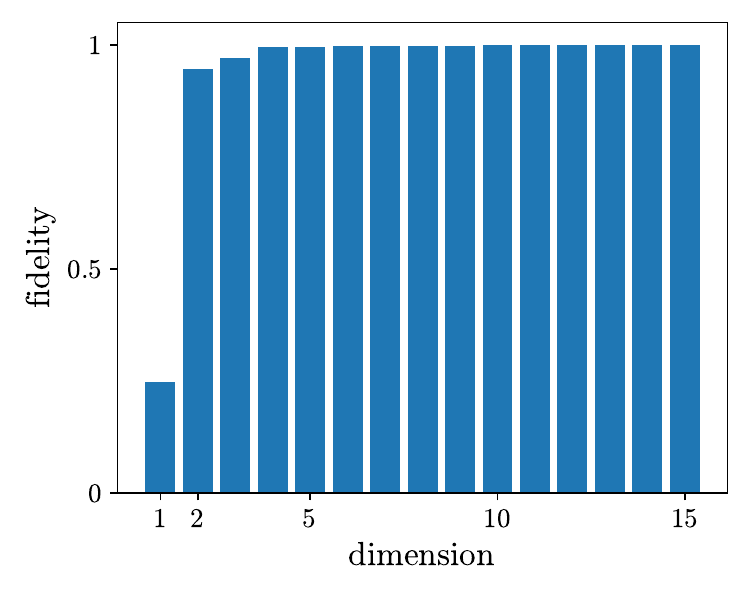}
\includegraphics[width=0.3\columnwidth]{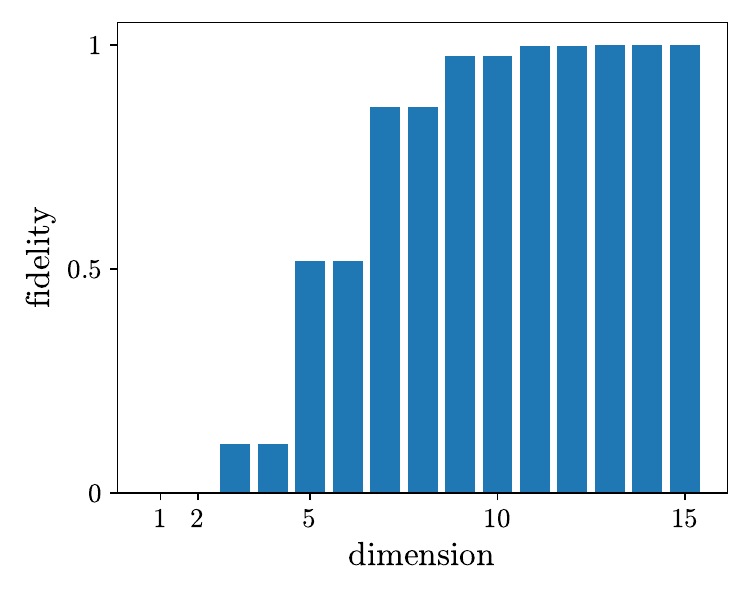}
\caption{
Left panel: Wigner function of the target even cat state. Middle panel: reconstruction fidelity versus the dimension of the subspace defined by the dominant eigenmodes of the Gram operator $G$. Right panel: reconstruction fidelity versus reconstruction dimension in the Fock basis. The Gram basis reaches high fidelity within a remarkably low-dimensional subspace: only three dominant modes are sufficient, whereas approximately ten Fock states are required for comparable reconstruction quality.
}
\label{Cat}
\end{figure}

\begin{figure}[ht]
\centering
\includegraphics[width=1\columnwidth]{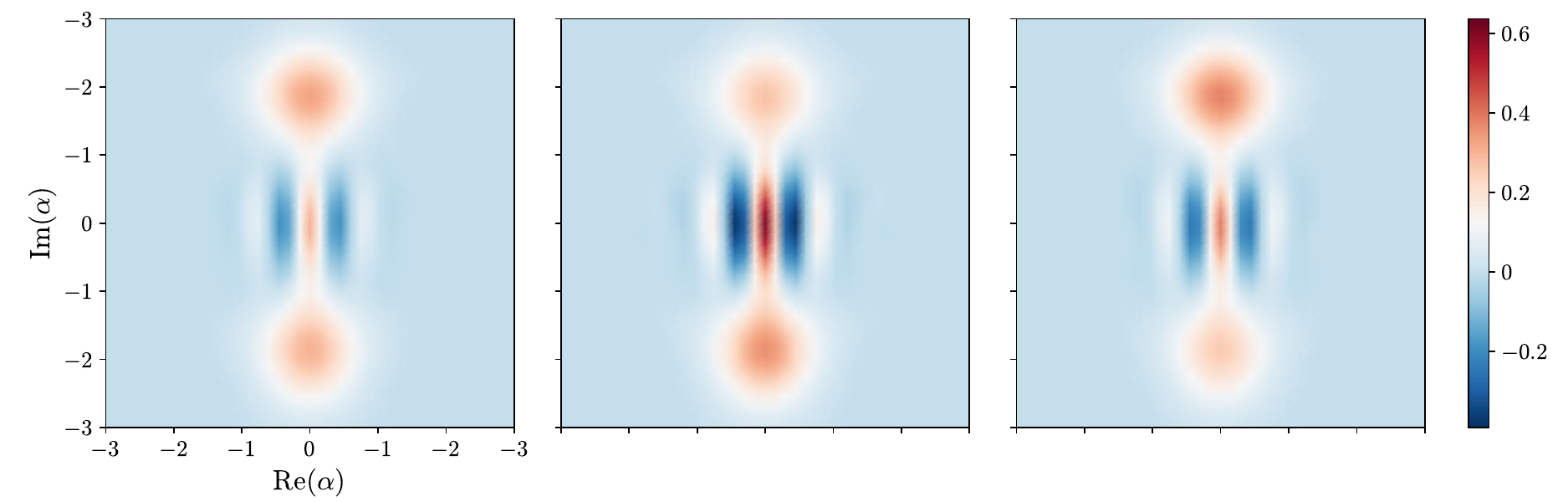}
\hfill
\includegraphics[width=1\columnwidth]
{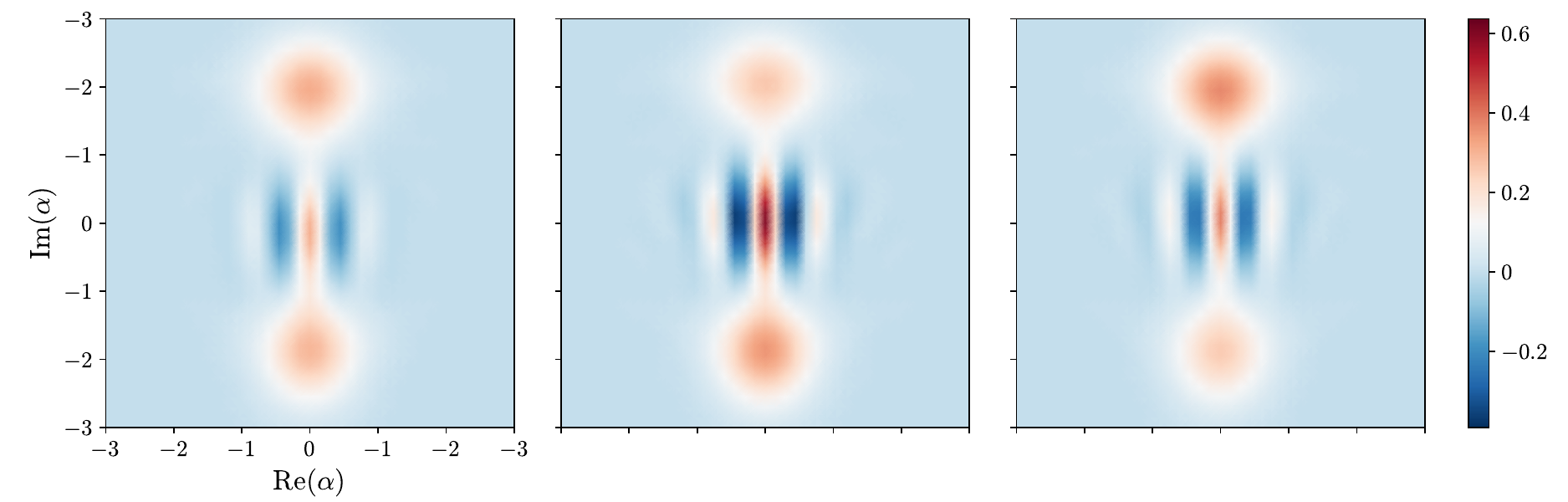}
\hfil
\includegraphics[width=1\columnwidth]
{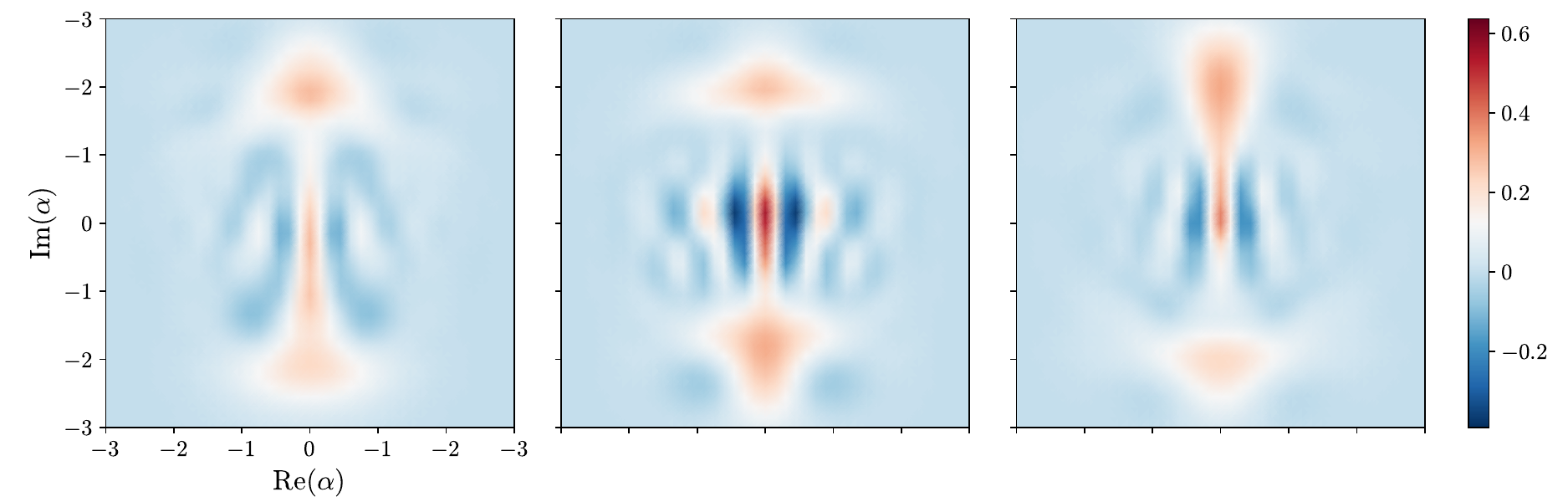}
\caption{
Repeated MaxLik reconstructions in subspaces spanned by the 2, 3, and 11 dominant Gram modes. Increasing dimensionality reveals finer structures but enhances sensitivity to statistical fluctuations.
}
\label{Rec}
\end{figure}

{\em Numerical simulations and discussion:}
Several numerical examples illustrate how finite sampling constrains the operational resolution of phase-space tomography with noisy data. As a representative example, we consider homodyne tomography of an even cat-like state similar to those investigated experimentally in Refs.~\cite{Ourjoumtsev2006,science24}. The simulated measurement comprises six homodyne phase cuts with 51 quadrature bins uniformly distributed over the interval $(-5,5).$  The target state is an even coherent-state superposition with amplitudes $\pm 2$.

Reconstruction is carried out in a Fock space truncated at $n=14$, followed by analysis in the modal basis defined by the Gram operator $G$. This representation makes it possible to compare reconstruction efficiency in a predetermined basis with the effective low-dimensional structure selected by the measurement itself. Statistical fluctuations are simulated through Poissonian noise in the measured counts.

Figure~\ref{Gram} shows the eigenvalue spectra of the Gram operator $G$ and the operator-space Gram matrix $Q$ for simulated homodyne tomography. The spectrum of $G$ determines the effective reconstruction bandwidth by identifying the experimentally accessible modes, whereas $Q$ quantifies the quadratic conditioning structure governing noise sensitivity in operator space.

Figure~\ref{Cat} compares the exact Wigner function of the target cat state with the reconstruction fidelity 
\begin{align}
F=|\langle\psi|\psi_d\rangle|^2 ,
\end{align}
evaluated as a function of the reconstruction dimension $d$. The middle panel shows reconstruction in the modal basis defined by the eigenvectors of $G$, whereas the right panel shows reconstruction in a predetermined Fock basis. 
The Gram basis provides an efficient measurement-adapted representation of the quantum state. The experimentally accessible structure of the cat state is captured already within a remarkably low-dimensional subspace of three dominant Gram modes, whereas reconstruction in a predetermined Fock basis requires substantially larger dimensions to achieve comparable fidelity.

\begin{figure}[ht]
\centering
\includegraphics[width=1\columnwidth]{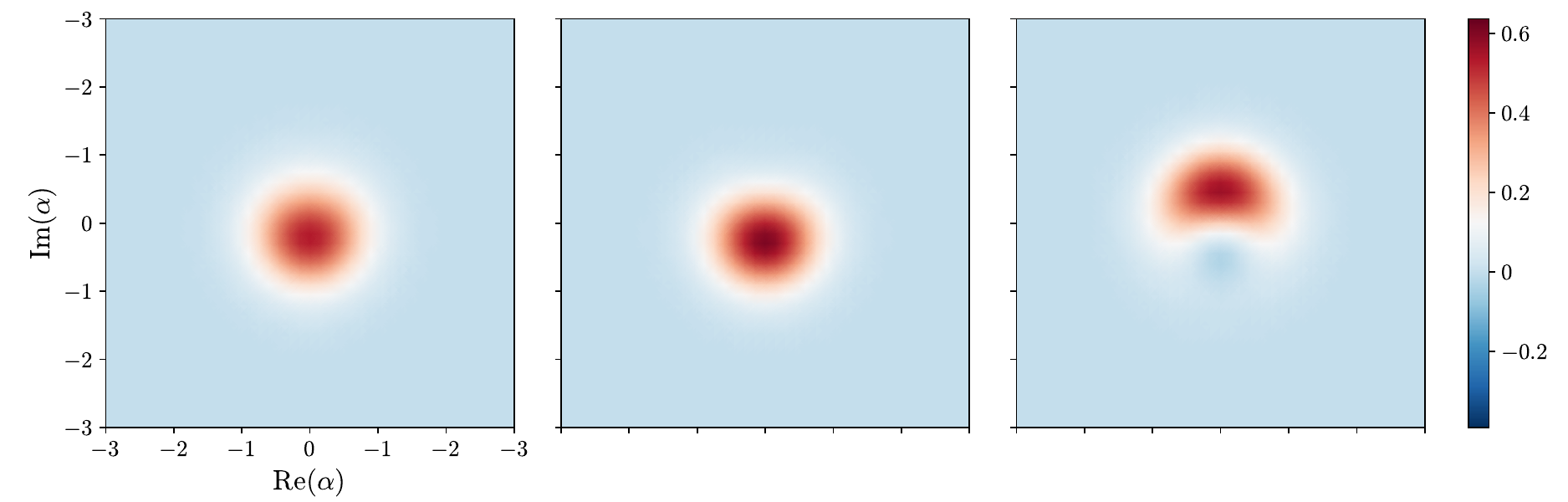}
\hfill
\includegraphics[width=1\columnwidth]
{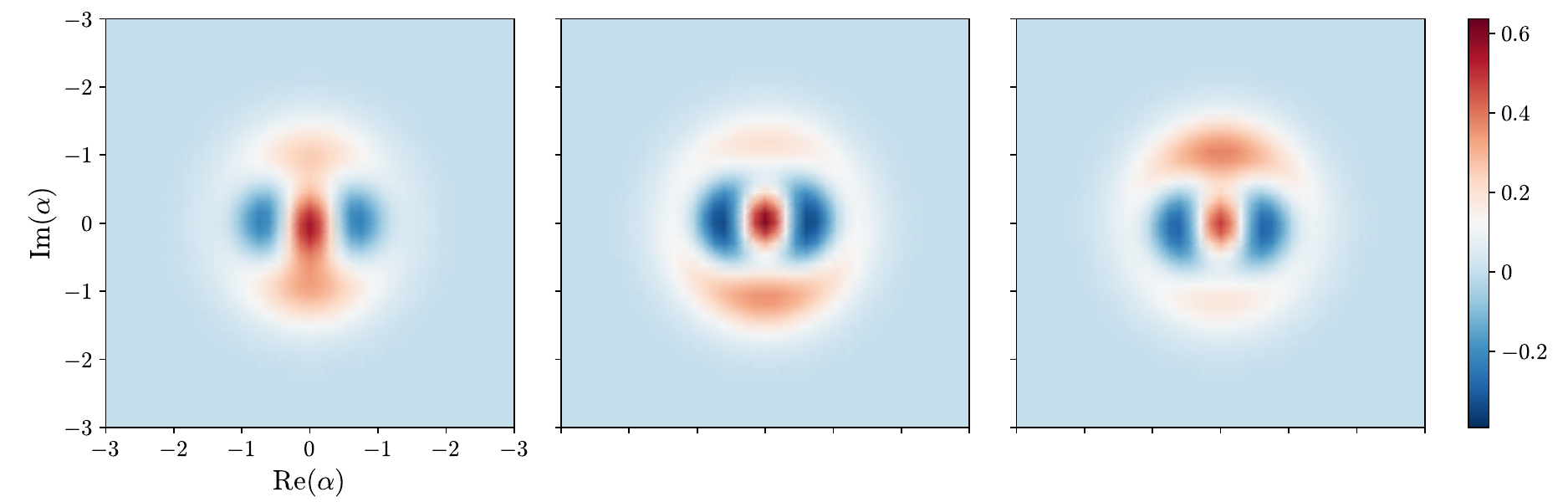}
\hfil
\includegraphics[width=1\columnwidth]
{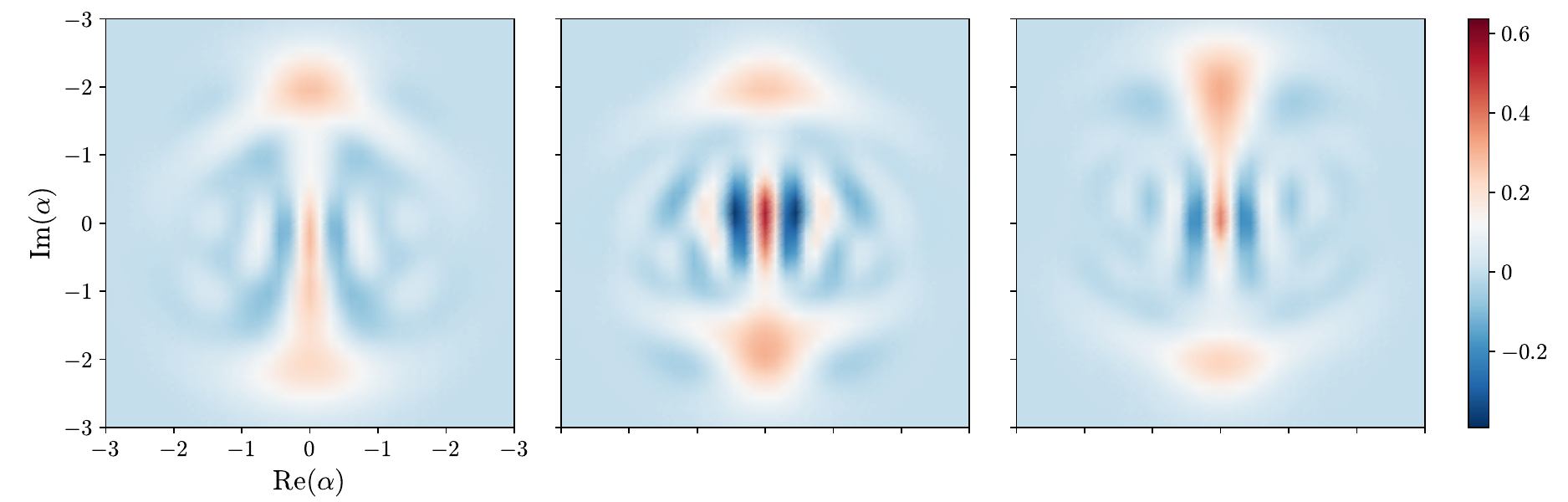}
\caption{
Repeated MaxLik reconstructions in Fock subspaces of dimensions 2, 3, and 11. Comparable structures require larger dimensions and exhibit stronger fluctuations than in the Gram basis.
}
\label{Rec_F}
\end{figure}

Figures \ref{Rec} and \ref{Rec_F} compare repeated MaxLik reconstructions in the measurement-adapted Gram basis and in a predetermined Fock basis. In the Gram basis, only a few dominant modes are sufficient to recover the experimentally accessible structure, whereas weakly sampled higher modes mainly increase noise sensitivity. In the Fock basis, comparable features appear only at substantially larger dimensions and are accompanied by pronounced fluctuations between repeated realizations due to the fitting of the noise.



\begin{figure}[ht]
\centering
\includegraphics[width=1\columnwidth]{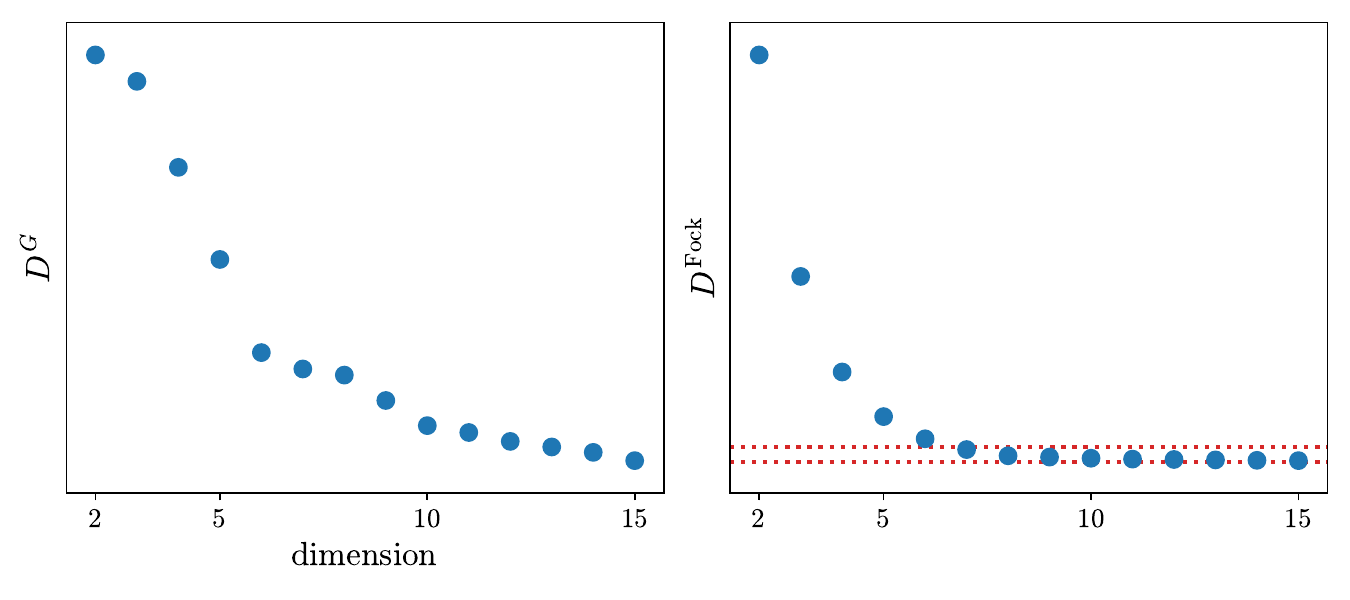}
\caption{
Deviance versus reconstruction dimension in the Gram eigenbasis (left) and in a predetermined Fock basis (right). The crossover visible in the Gram basis identifies the effective measurement-supported reconstruction bandwidth. Red dashed lines indicate the range of deviance values obtained in the Gram basis.
}

\label{dev}
\end{figure}


Figure \ref{dev} provides an operational signature of the effective reconstruction bandwidth. In the Gram basis  on the left panel the deviance exhibits a clear crossover between dominant    measurement-supported modes and weakly sampled components. This is  consistent with the stability observed in Fig.\ref{Rec},  where the essential structure of the target state is already recovered within a low-dimensional Gram subspace.
In contrast, the deviance in the  Fock basis (right panel)  shows an almost monotonic decrease, indicating that additional parameters improve the fit without providing a comparable increase in experimentally supported information.

The measurement-adapted Gram basis makes the transition between strongly and weakly supported reconstruction modes directly visible in the likelihood statistics. In predetermined bases, this transition is generally less apparent and often requires additional regularization or model-selection procedures.

Although related ideas have appeared in earlier developments of MaxLik tomography~\cite{Hradil2004,Hradil_06Biased,Rehacek_2008Tomography}, the explicit interpretation of the operator $G$ as a Gram operator defining the operational reconstruction bandwidth, together with the resulting measurement-adapted reconstruction strategy, has not been formulated previously. The framework developed here is immediately applicable tool for contemporary experiments probing highly structured states in large Hilbert spaces.

Since the boundary between resolved and unresolved features is rarely sharp, maximizing the likelihood alone does not necessarily provide the most reliable description. Enlarging the reconstruction space may improve the fit while simultaneously amplifying statistical fluctuations associated with weakly sampled modes. In such situations, more sophisticated statistical approaches, including penalized likelihood, information criteria, maximum-entropy methods, or error-region analysis, may provide a more rigorous assessment of experimentally inferred features \cite{Englert2025}. From this perspective, the Gram spectrum provides a natural zeroth-order diagnostic by identifying the statistically supported subspace of Hilbert space before more sophisticated inference procedures, such as regularization or error-region analysis, are invoked.

{\em Summary: }  We have shown that the reliability of features inferred from quantum tomography is fundamentally limited by the resolving power of the measurement, quantified by the Gram (sampling) operator associated with MaxLik estimation. The resulting picture establishes a direct analogy with classical imaging, where structures appearing near the resolution limit of an instrument must be interpreted with care.

Reconstruction in the Gram eigenbasis emerges as an efficient measurement-adapted compression of the tomographic problem, revealing the experimentally accessible bandwidth while quantifying the role of conditioning and statistical fluctuations. From this perspective, the central question is not whether a reconstruction algorithm can generate a particular feature, but whether the measurement itself possesses sufficient resolving power to support it. The framework developed here provides an operational criterion for addressing this question. 
In Eddington's spirit, the Gram operator may be regarded as a quantum fishing net. The reconstruction is naturally restricted to the space explored by typical measurement outcomes, since common events occur commonly whereas rare events occur rarely.

\begin{acknowledgments}
We acknowledge the support of the  Czech  Science Foundation   under the grant agreement 26-22242J.
\end{acknowledgments}
\bibliography{Masterbib}
\clearpage
\onecolumngrid
\input{Supplementary_input}

\end{document}

%% file: Supplementary_input.tex
\clearpage
\onecolumngrid

\section*{Supplementary Material}

\maketitle

\onecolumngrid


\setcounter{equation}{0}
\setcounter{figure}{0}
\setcounter{table}{0}

\renewcommand{\theequation}{S\arabic{equation}}
\renewcommand{\thefigure}{S\arabic{figure}}
\renewcommand{\thetable}{S\arabic{table}}


\section{Finite frame theory}

It is intriguing to note that elegant  unifying mathematical formulation of inversion problems can be given in terms of finite frame theory \cite{Waldron2018}, which  formulates signal expansion and decomposition in terms of non-orthogonal and overcomplete bases.

{\it Finite frame theory in state space: }  Let us first show that the operator $G$ is unitarily equivalent to the Gram matrix associated with projective measurements described by POVM elements
\begin{align}
\Pi_i = |y_i\rangle\langle y_i|,
\end{align}
for which
\begin{align}
G_{ij}=\langle y_i|y_j\rangle .
\end{align}
This follows directly for any set of linearly independent vectors $|y_i\rangle$ forming a generally nonorthogonal basis by considering superpositions
\begin{align}
|g\rangle=\sum_i a_i |y_i\rangle .
\end{align}

The role of the Gram operator in quantum tomography was discussed in the main text. Here we formulate the same structure within finite frame theory, where $G$ plays the role of an effective frame operator governing the stability of the reconstruction.

The associated dual frame is defined by
\begin{align}
|\tilde y_i\rangle = G^{-1}|y_i\rangle ,
\end{align}
which yields the reconstruction formula
\begin{align}
|\psi\rangle
=
\sum_i
\langle \tilde y_i|\psi\rangle |y_i\rangle
=
\sum_i
\langle y_i|\psi\rangle |\tilde y_i\rangle .
\end{align}

From this perspective, MaxLik estimation may be interpreted as an implicit regularization of the inversion of the tomographic frame. The nonlinear operator
\begin{align}
R(\rho)=\sum_i \frac{f_i}{p_i(\rho)}\,\Pi_i
\end{align}
adaptively reweights the measurement operators according to the observed data. The reconstruction can therefore be viewed as employing a data-dependent dual frame which suppresses poorly sampled modes while simultaneously preserving the physical constraint of positive semidefiniteness. In this sense, MaxLik tomography performs a constrained inversion of the tomographic frame that naturally stabilizes the reconstruction in the presence of incomplete or noisy sampling.


{\it Finite frame theory in operator space: }   A closely related formulation arises in the Hilbert--Schmidt operator space associated with linear inversion tomography. The operator frame is generated by the projectors
\begin{align}
\Pi_i = |y_i\rangle\langle y_i|.
\end{align}

In vectorized form these operators define a frame whose frame operator acts on a generic operator $A$ according to
\begin{align}
S(A)=\sum_i \mathrm{Tr}(\Pi_i A)\,\Pi_i .
\end{align}
The corresponding Gram matrix is
\begin{align}
Q_{ij}
=
\mathrm{Tr}(\Pi_i\Pi_j)
=
|\langle y_i|y_j\rangle|^2 .
\end{align}
Linear inversion reconstructs the density operator using the dual frame
\begin{align}
\rho
=
\sum_i
\mathrm{Tr}(\rho \Pi_i)\,
\tilde\Pi_i ,
\end{align}
where the dual-frame elements are defined by
\begin{align}
\tilde\Pi_i
=
S^{-1}(\Pi_i).
\end{align}

The connection with modal decomposition and singular-value regularization becomes explicit when the density operator is expanded in the nonorthogonal basis generated by the projectors $|y_i\rangle$,
\begin{align}
\rho
=
\sum_{jk}
\rho_{jk}
|y_j\rangle\langle y_k|.
\end{align}

The measured probabilities satisfy
\begin{align}
p_i
=
\langle y_i|\rho|y_i\rangle
=
\sum_{jk}
G_{ij}\rho_{jk}G_{ki},
\end{align}
Since Gram matrix  $G$ is Hermitian and positive semidefinite, it admits the spectral (and singular value) decomposition
\begin{align}
G = U \Lambda U^\dagger,
\end{align}
with $\Lambda = \mathrm{diag}(\lambda_1,\dots,\lambda_r)$.
Transforming to the modal basis defined by $U$, with $\tilde{\rho} = U^\dagger \rho U$, one finds
\begin{align}
G \rho G = U \Lambda \tilde{\rho} \Lambda U^\dagger .
\end{align}
In this representation, each matrix element is weighted as
\begin{align}
(\Lambda \tilde{\rho} \Lambda)_{kl}
= \lambda _k \lambda_l \tilde{\rho}_{kl}.
\end{align}

Coherences between weakly sampled modes are suppressed quadratically through the products $\lambda_k \lambda_l$.
The frame-theoretic formulation therefore makes explicit that both MaxLik tomography and linear inversion are governed by the same underlying modal structure determined by the Gram operator and its singular-value spectrum.
The two Gram structures  encode complementary aspects of tomography: the Gram operator in Hilbert space specifies the effective reconstruction bandwidth, while the Gram matrix in operator space determines the conditioning and noise sensitivity of the inversion.

In mathematical terms  operator-space Gram matrix is related to the Hilbert-space Gram operator through the Hadamard product
\begin{align}
Q=G\circ G^*,
\end{align}
which reflects the fact that operator reconstruction inherits the measurement bandwidth simultaneously through bra and ket components. Consequently, conditioning in operator space is governed by a quadratic sensitivity structure relative to the underlying state-space sampling.

{\em Optimization of conditional likelihood}

The conditional likelihood can be written as
\begin{equation}
\mathcal{L}_{\mathrm{cond}}(\rho)
=
\prod_i
\left(
\frac{p_i(\rho)}
     {G(\rho)}
\right)^{n_i},
\end{equation}
where
\begin{equation}
p_i(\rho)=\mathrm{Tr}(\rho\Pi_i), \quad G(\rho)
=
\mathrm{Tr}(\rho\Gamma),
\qquad
\Gamma=\sum_{i\in S}\Pi_i .
\end{equation}
The corresponding log-likelihood reads
\begin{equation}\log 
{\mathcal L}_{\mathrm{cond}}(\rho)
=
\sum_i n_i \log p_i(\rho)
-
N \log G(\rho),
\qquad
N=\sum_i n_i .
\end{equation}

The first term coincides with the standard maximum-likelihood objective and is concave over the convex set of density operators. The second term introduces a nonlinear dependence through the normalization functional \(G(\rho)\), rendering the overall optimization problem non-convex. A useful observation is that for a fixed value
$ G(\rho)=g,  $
the second term becomes constant and the optimization reduces to the standard maximum-likelihood problem on this hyper-surface
\begin{equation}
\max_{\rho}
\sum_i n_i \log p_i(\rho)
\quad
\text{subject to}
\quad
G(\rho)=g .
\end{equation}
Hence, the conditional likelihood optimization may be viewed as a family of conventional convex maximum-likelihood problems parameterized by the value of \(g\).
  Since \(\Gamma\) is a positive operator, the normalization functional is bounded by the extremal eigenvalues of \(\Gamma\),
\begin{equation}
\lambda_{\min}(\Gamma)
\le
G(\rho)
\le
\lambda_{\max}(\Gamma),
\end{equation}
for all density operators \(\rho\). Therefore, the optimization effectively explores a one-parameter family of convex likelihood maximizations over the interval
\begin{equation}
g \in
\big[
\lambda_{\min}(\Gamma),
\lambda_{\max}(\Gamma)
\big].
\end{equation}

From this perspective, the non-convexity of the conditional likelihood originates solely from the additional optimization over the admissible values of the normalization functional \(G(\rho)\), whereas each fixed-\(g\) subproblem remains a standard convex maximum-likelihood estimation.  The functional \(G(\rho)\) is determined by the Gram operator \(\Gamma\), which characterizes the measurement subspace associated with the observed outcomes. Consequently, the conditional likelihood may be interpreted as an optimization not only over the quantum state \(\rho\), but also over the effective measurement normalization induced by the corresponding Gram operator.